# Specific interactions of peripheral membrane proteins with lipids: what can molecular simulations show us?

Andreas H. Larsen[1,2], Laura H. John[1], Mark S.P. Sansom[1], Robin A. Corey[1*]

[1] *Department of Biochemistry, University of Oxford*
[2] current address: *Department of Neuroscience, Faculty of Health and Medical Sciences (SUND), University of Copenhagen*

* to whom correspondence should be addressed: robin.corey@bioch.ox.ac.uk

## Abstract

Peripheral membrane proteins can reversibly and specifically bind to biological membranes to carry out functions such as cell signalling, enzymatic activity, or membrane remodelling. Structures of these proteins and of their lipid-binding domains are typically solved in a soluble form, sometimes with a lipid or lipid headgroup at the binding site. To provide a detailed molecular view of peripheral membrane protein interactions with the membrane, computational methods such as molecular dynamics (MD) simulations can be applied. Here, we outline recent attempts to characterise these binding interactions, focusing on both intracellular proteins such as PIP-binding domains, and on extracellular proteins such as glycolipid-binding bacterial exotoxins. We compare methods to identify and analyse lipid binding sites from simulation data. We highlight recent work characterising the energetics of these interactions using free energy calculations. We describe how improvements in methodologies and computing power will help MD simulations to continue to contribute to this field in the future.

## Introduction

Peripheral membrane proteins (PMPs) bind reversibly to the surface of specific biological membranes, and thus can exist in both a soluble and membrane-bound state. Typically, membrane binding events are transient, and involve the PMP covalently (via a lipid anchor) and/or non-covalently interacting with the surface of the membrane. This ability to switch between a highly mobile soluble state and a specifically targeted membrane-bound state is often central to the PMP function. In particular, PMPs often interact with specific lipids in the membrane, making understating these interactions crucial to understanding the biology of PMPs.

PMPs are vital in a range of cellular processes, including signalling, trafficking, apoptosis and immunity. This makes PMPs attractive as drug targets, with around 30 human PMPs currently being targeted (1,2). In particular, the lipid-binding domains of PMPs are promising targets for future therapeutics (3), as the membrane binding process is potentially open to modulation. Moreover, many PMPs are multi-domain proteins, with one or more domains driving the binding to the membrane, and other domains performing the downstream function of the protein, e.g. acting as enzyme (Figure 1). This means that PMPs can potentially play important roles in biotechnology or medicine, such as drug delivery (4). For all of these aspects of PMPs understanding the membrane binding process, which is usually driven by lipid interactions, is crucial.

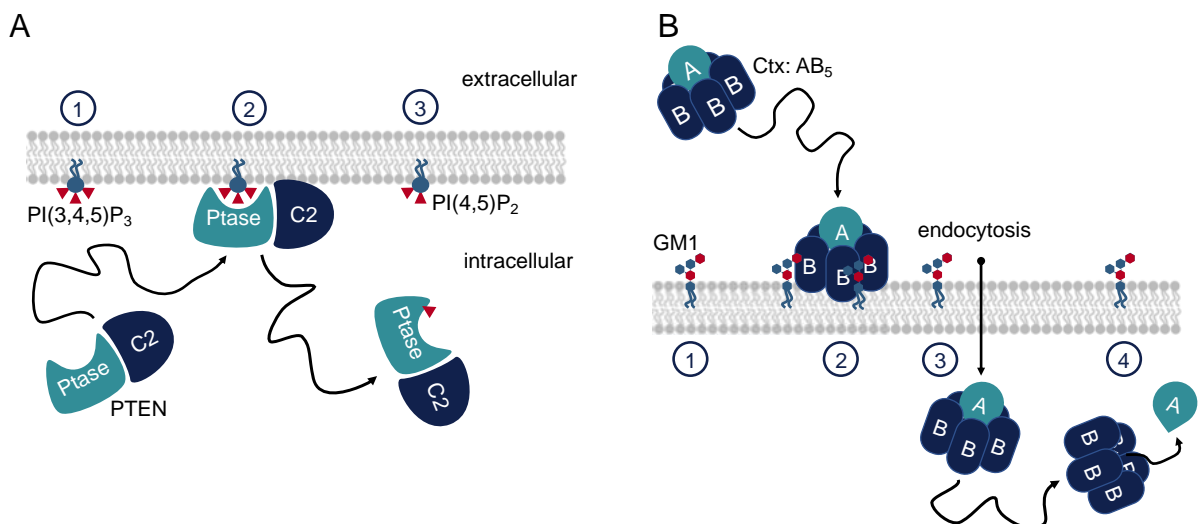

***Figure 1. Two examples of PMP action.*** *(A) Phosphatase and tensin homologue (PTEN) is a PMP that consists of an enzymatic domain, the phosphatase domain (Ptase), and a lipid-binding C2 domain. PTEN binds to the inner leaflet of eukaryotic membranes, where Ptase dephosphorylates $PI(3,4,5)P_3$ to form $PI(4,5)P_2$. This process is part of a signalling pathway leading to apoptosis. (B) Cholera Toxin (Ctx) is also a PMP with an enzymatic domain, the A subunit, and a lipid-binding domain consisting of five B subunits. This $B_5$ ring binds to the ganglioside GM1 and following the binding event, the complex is translocated into the cell by endocytosis. After translocation, the A subunit separates from the Ctx complex and carries out its cytotoxic mechanism.*

There are many ways in which PMPs can target specific lipids. In many cases lipid recognition is through direct electrostatic interactions, often between basic residues on the PMPs and lipids with anionic headgroups (Figure 2A). These interactions can be very strong and can involve lipids such as the monovalent phosphatidylserine (PS) or multivalent phosphatidylinositol phosphate (PIP) lipids e.g. $PIP_2$ and $PIP_3$, both present in the inner leaflet of eukaryotic cell membranes. Alternatively, for PMPs which bind extracellularly, gangliosides containing negatively-charged sialic acids are often important (5–7). Electrostatic interactions can be mediated via positive ions (typically calcium) that bridge negatively-charged patches on the PMP with negative lipids (8,9) (Figure 2B). Alternative forms of interaction exist, such as insertion of hydrophobic loops (10–16) or hydrophobic helices (17) into the membrane, cation-π interactions between aromatic residues and choline headgroups (18,19), or covalent attachment to a lipid whose hydrophobic tail is inserted into the lipid bilayer (20) (Figure 2).

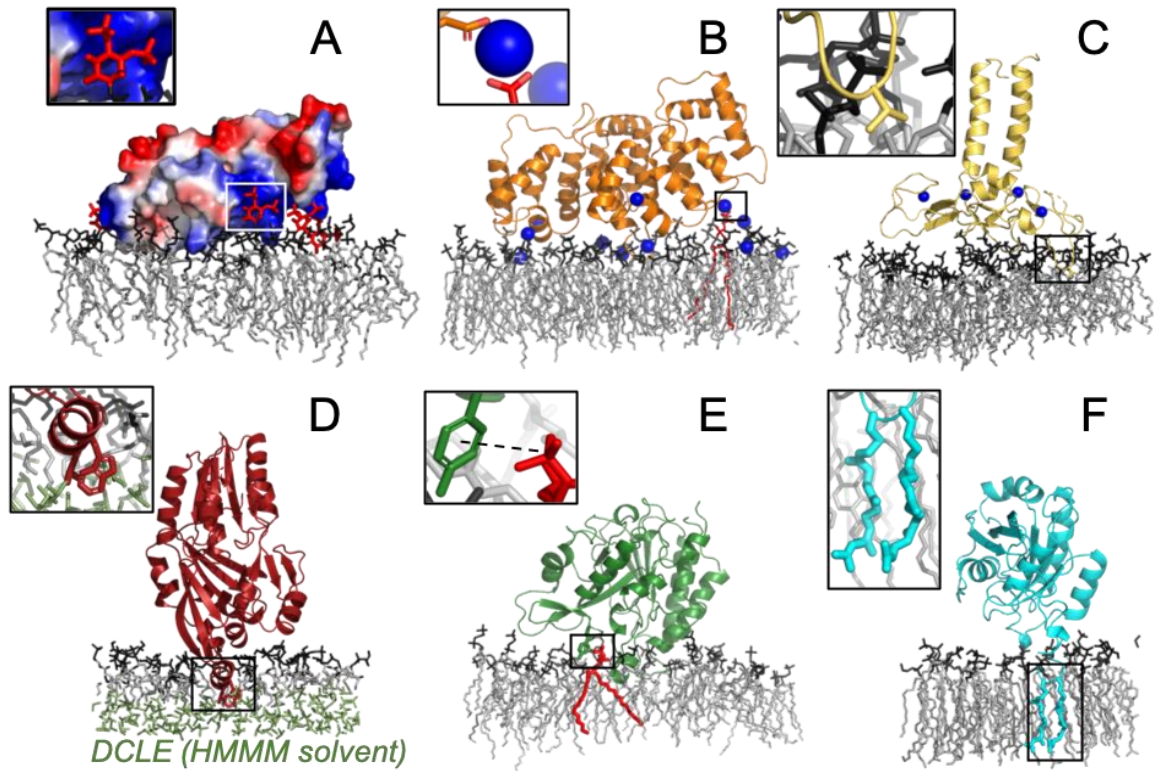

**Figure 2. Different binding modes for PMPs.** *(A) Direct, electrostatic binding. The C2 domain of RIM2 has a cationic groove (blue) that binds to anionic $PIP_2$ (21). (B) Ion-promoted electrostatic binding. Annexin A2 binds to an anionic membrane bridged by calcium ions. Insert shows a $Ca^{2+}$ ion bridging a negatively charged phosphatidylserine headgroup (red) and an aspartic acid side chain (D166) from Annexin A2 (9). (C) Loop insertion. The FYVE domain of EEA1 (zinc ions from the four zinc fingers in blue) has a partly hydrophobic loop (V44 shown in inset) which plays part in the membrane-binding (16). (D) Helix insertion. FakB1 inserts a helix in the membrane to expose hydrophobic residues of that helix to the hydrophobic membrane interior (22). Note that HMMM was used in this study (see main text), so short lipids were used together with a hydrophobic solvent, DCLE (green), to represent the rest of the lipid tails. (E) Cation-π interaction. Membrane-binding mediated by cation-π interaction between residues Y88 of PI-PLC (green) and a DMPC headgroup (red) (19). (F) Binding by lipidation. Rab5 binds to membranes via its two geranylgeranyl lipid anchors (23). Coordinates for the figure were kindly provided by the authors of the respective papers.*

The soluble states of lipid recognition domains of PMPs are generally straightforward to determine using X-ray crystallography or solution NMR. It can be challenging, however, to obtain a detailed atomic resolution description of the membrane bound state of PMPs, in part because of the difficulty in stabilising membrane-bound states for structural analysis. In addition, there can be considerable complexity in the interaction which can make the exact state difficult to recreate experimentally. This includes such factors as the binding of ions (24), the macrodipole of the PMP (25), and the structural and chemical complexity of biological membranes (26).

In this review, we present how molecular dynamics (MD) simulations can be used to predict the membrane-bound state of PMPs. In the first part, we will focus on the identification of specific protein-lipid interactions, i.e., the binding of a particular lipid species to a residue or cluster of residues on the surface of a PMP. In the second part, we will review methods to calculate binding affinities and avidities for such binding events. Finally, we will outline future trends and discuss challenges in computational and experimental analyses of PMPs.

## Biological background

In this section, we will give a brief overview of the biological roles of some key PMPs that bind to the cytoplasmic plasma membrane, either inside or outside of the cell. This is far from being a comprehensive list, but rather focuses on examples which have been investigated with MD and demonstrate specific lipid-binding events.

PMPs can be divided into two broad groups, intracellular and extracellular. These groups are split both in terms of their biological role and the type of lipids they bind, owing to the strongly asymmetric nature of the plasma membrane (27,28). Here, we will mostly focus on intracellular PMPs which bind PIP lipids, and extracellular PMPs which bind to gangliosides. The former group typically carry out a role necessary to the functioning of the cell, whereas the latter group includes secreted bacterial exotoxins which bind and attach to the host membrane.

### Biological roles of intracellular peripheral membrane proteins

Many cytoplasmic PMPs in eukaryotic cells use polybasic binding domains to bind negatively charged phospholipids, which are enriched in the inner cytoplasmic leaflet (29). These include phosphatidylserines (PS), which are the most abundant anionic phospholipids in the inner leaflet. Although less abundant, the most important PMP-recruiting anionic lipids are probably phosphatidylinositols, PIs (30–32). Importantly, PIs can be phosphorylated, which is utilized extensively in cell signalling processes, with the phosphorylation or dephosphorylation often carried out by PMPs. One example is the phosphatase and tension homologue (PTEN), which dephosphorylates $PI(3,4,5)P_3$ to $PI(4,5)P_2$. This is the first step in a pathway leading to apoptosis, so malfunctioning of PTEN can cause tumours and several mutations of PTEN are thus linked to cancer (33). PTEN consists of a phosphatase domain, that dephosphorylates $PIP_3$, and a lipid-binding C2 domain (Figure 1A). C2 is a large family of lipid binding domains that exist in different proteins across species. Another abundant PI binding family is that of PH domains, which is one of the largest families of lipid-binding domains, existing in more than 250 human proteins (34). The PH domains vary in sequence, but the structure is conserved across protein families. Many PH domains show high selectivity to phosphorylated PI, thereby enabling spatial and temporal docking of their parent protein (35). An example of a PH domain-containing protein is the General Receptor for Phosphoinositides 1 (Grp1), which binds to $PIP_3$ with unusually high selectivity and affinity (36–38). Grp1 is a GDP/GTP exchange factor that catalyses the activation of ADP ribosylation factor proteins, involved in actin rearrangement, endo- and exocytosis, membrane budding and trafficking (39). Like in most PMPs the membrane-anchoring domain is not the same as the catalytic domain, which in the case of Grp1 is Sec7.

Other examples of lipid-binding domains are FYVE, PX, ENTH, CALM, PDZ and PTB domains (40), which also exist in many different PMPs. The FYVE domain, for example, is part of Early Endosome Antigen 1 (EEA1), which plays a key role in recruiting early endosomes, and FYVE acts as a lipid anchor for EEA1 (Figure 2C). The structure of many of these lipid-binding domains is well described. However, their lipid-binding properties are, in many cases, still controversial. This is the case even for extensively studied domains like PH: while a genome wide study using 33 yeast PH domains showed that most *Saccharomyces cerevisiae* PH domains bind PIPs very weakly or not at all (41), a recent study using 67 human PH domain-containing proteins found that 54% of the proteins had significant affinity for PIPs (34).

## Biological roles of extracellular peripheral membrane proteins

As well as PMPs which bind to the plasma membrane from the cytoplasm, there are those which bind the extracellular face of the membrane. A notable category of these are bacterial exotoxins, which are secreted by certain pathogenic bacteria to target host membranes. Exotoxins are typified by having a high potency, with just a few copies able to cause severe damage to their target cells. Many exotoxins either disrupt or locally permeabilize the membrane or change the environment around the cell to be favourable to the bacterium.

There are several classes of bacterial exotoxin. One well-studied example is the $AB_5$ exotoxins, which bind to specific glycolipids on the exterior of the cell membrane. A prominent example of this is cholera toxin (Ctx), secreted by the *Vibrio cholerae* bacterium. Ctx has a $B_5$ component, comprising five B subunits arranged in a ring shape, and an A subunit which sits on top of the ring (42) (Figure 1B). The $B_5$ ring is responsible for recognising and binding specific receptors in the host membrane. For Ctx, the primary receptor is the ganglioside GM1 (43). In addition, secondary binding sites are also present on the $B_5$ ring for fucosylated Lewis antigens on the membrane of epithelial cells (44,45). Another prominent example of $AB_5$ exotoxins includes the structurally similar *Escherichia coli* heat-labile enterotoxin (LT), which also binds GM1 (46) as well as other gangliosides (47). This too has secondary binding sites, this time for A-type blood antigens (44). A third example is the family of Shiga toxins (SHT) from *Shigella dysenteriae* and some *E. coli* strains, which binds to globosides, which are uncharged glycosphingolipids (48).

Together, bacteria which secrete $AB_5$ account for over a billion infections and a million deaths per year. In addition, the $B_5$ form of these toxins have value of their own, in both therapeutics (49) and biotechnology (50).

## Section 1: Use of molecular dynamics to identify lipid binding sites

## Identifying PMP orientations and lipid binding sites

To understand the biological activity of a given PMP, it is essential to know how it interacts with the membrane, including its orientation relative to the lipid bilayer. For many PMPs, identifying specific interactions with lipids in the membrane is an important component of this. One way to investigate specific interactions is through computational analysis, such as with MD. In Figure 3A, we provide an overview of different MD-based computational methods commonly used to determine the orientation of PMPs, and the binding sites for specific lipids, and discuss each in turn below.

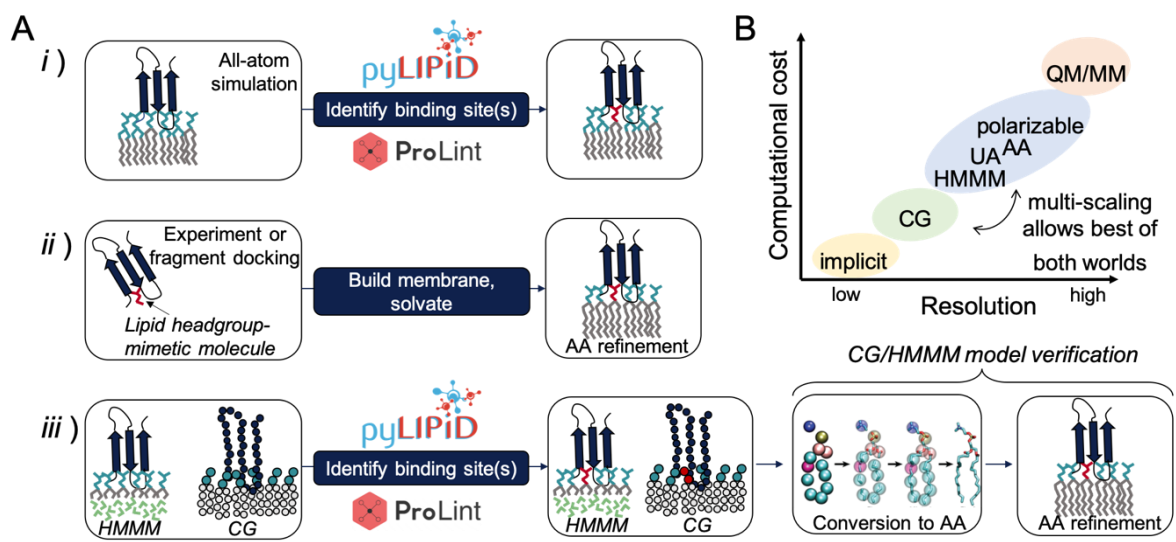

**Figure 3. Strategies for simulating PMP binding with molecular dynamics simulations.** *(A-i) All-atom (AA) simulation is performed, and binding sites are identified, e.g. with PyLipID or ProLint. (A-ii) Experiment or fragment docking simulations are used to identify binding sites of a PMP in solution. The fragments are, in this context, lipid headgroup mimicking molecules. The protein can then be solvated, and fragments replaced by whole lipids. (A-iii) The Highly Mobile Membrane Mimetic (HMMM) model or coarse grained (CG) simulation are used to sample binding events of PMP binding to lipid membrane, and binding sites are subsequently identified. To verify the simulations, the binding sites are converted to all-atom resolution and refined in an AA simulation. The panel shows the modular conversion of palmitoyl-oleoyl-glycero-phosphatidylcholine (POPC) from CG to AA resolution using the program CG2AT (51); panel adapted from https://github.com/owenvickery/cg2at. (B) Overview of different simulation approaches and how they qualitatively compare with respect to resolution and computational cost. Implicit methods utilize mean force description of membrane and/or solvent to speed up the calculations. CG, HMMM, UA and AA are all described in the main text. Quantum mechanics/molecular mechanics (QM/MM) simulations are precise and can be used to model formation and breaking of covalent bonds. QM/MM simulations are, however, (yet) too computationally expensive for simulations of PMPs binding to lipid membranes, but included for completeness.*

*Atomistic molecular dynamics simulations*

Atomistic simulations are a well-established and widely used flavour of MD. These simulations explicitly model all atoms (AA), including hydrogens, in a system as point charges with no electronic structure (Figure 3B). A variant is united-atom (UA) force fields, where certain hydrogens in the protein are incorporated into the particles of associated heavy atoms, so that e.g. -$CH_3$ might be represented as a single particle. UA force fields are computationally less costly but with slightly lower resolution, although they are not necessarily less accurate than AA force fields (52).

There are many choices necessary to make when setting up atomistic MD simulations of PMPs. A key decision is the choice of force field used to describe the protein, solvent and lipids, with many options available, including different UA and AA force fields. In addition, in certain force fields the protein and/or water can be polarized, for a potentially more accurate view of certain molecular phenomena (53). Here, we will mostly focus on AA force fields, e.g. CHARMM36m (54), Amber (55), and OPLS-AA (56). For a more detailed discussion on these choices, we recommend referring to (57–59).

The use of AA can in principle be very simple, as all that is needed is to place the PMP near a desired membrane and run an MD simulation e.g. as done for the yeast oxysterol binding protein (Osh4) (11). Unfortunately, due to the long timescales needed to sample PMP binding to membranes, AA MD may be computationally too expensive even for small PMPs or isolated PMP domains. There are solutions available, such as using a massively parallel supercomputer, for example ANTON, which has been used to simulate Bruton's tyrosine kinase binding to a $PIP_3$ containing membrane (60). Alternatively, crowd-sourcing computational resource is another solution to generate enough sampling data, such as shown using Folding@home to model protein kinase C binding to the membrane (61). However, the high cost of these simulations is still prohibitive to many research groups, especially if multiple conformations or membrane lipid compositions are to be investigated.

Extensive simulations are not necessarily essential if only pose-generation, and not extensive sampling of the binding landscape, is desired. If the initial hurdle of generating the bound state can be overcome through either interpretation of experimental data or by a more approximative computational simulation, then AA can be applied to provide a high-resolution view of binding with a potentially much lower computational cost (Figure 3A).

*Use of experimental information to seed AA MD simulations*

A well-established method of generating an initial pose for AA MD is through integration of experimental data. A prominent example of this is with structural data, for instance where a fragment of a bound lipid has been co-crystallized with a PMP. This has been applied to $AB_5$ exotoxins such as Ctx (Figure 1B), which has a structurally well-defined GM1 binding pocket. In one computational study, the authors placed the toxin above a membrane and manually replaced the structurally-resolved sugars with full GM1 molecules from the membrane, to seed an MD simulation (62). Alternatively, if the PMP binds to an integral membrane protein as well as forming specific lipid interactions, then structural information of the complex can be used to orient the PMP, as is seen for the structure of bacterial SecA bound to SecYEG (63). From this input state, it is straightforward to setup subsequent MD simulations to identify lipid interactions (64). Of course, this approach is reliant on being able to resolve a structure of the PMP bound to an integral protein, which can be non-trivial.

Another technique which is often used to coordinate the PMP in relation to a specific lipid binding partner is electron paramagnetic resonance (EPR). Early examples of this include work incorporating EPR to model the C2 domain of cytosolic phospholipase A2 on a PC membrane (65). Docking information from EPR has also been applied in this manner to the C2 domain of Protein Kinase C$\alpha$ to PS and $PIP_2$ (66) and the Grp1 PH domain interactions with $PIP_3$ (67).

In theory, a range of methods can be combined to get a prior idea of the binding mode. For instance, Ohkubo and Tajkhorshid used both X-ray crystallography, NMR and fluorescence data to orient the GLA domain on a membrane. The GLA domain is a common membrane-anchoring domain of different PMPs. The group then used steered MD to pull the domain into the membrane and generate a bound state (68). A similar approach was also used for looking at how protein kinase C interacts with the membrane, with the 200 ns steered MD trajectory used to seed subsequent Folding@home simulations (61). Therefore, if the binding interface is already known, use of steered MD (69) can be a powerful way of sampling membrane binding in AA relatively quickly.

*If experimental data is unavailable: docking of lipid head fragments*

In the case that structural data is not available, or a different bound group is desired, a similar effect can be achieved using computational docking, where a soluble mimic of a lipid headgroup is docked onto the PMP binding pocket. Whilst not necessarily as robust as structural analysis, this approach is able to form a solid basis for downstream analysis using MD. This approach has, e.g., been applied to investigate the PIP binding sites of Smurf1-C2 (70), as well as interactions of the *E. coli* heat-labile enterotoxin with the blood group A pentasaccharide (44). The high throughput nature of the approach also makes docking of other molecules, such as potential therapeutic drugs, to the lipid binding sites feasible, as shown with an expansive analysis of alkaloid binding to Ctx (71).

*Accelerating lipid diffusion by the highly mobile membrane mimetic model*

An appealing method to achieve more effective sampling in AA simulations, is the highly mobile membrane mimetic (HMMM) model (72,73). In this approach, most of the volume normally filled by the lipid tails is replaced by an organic solvent, whilst the lipids are represented by their short-chain homologues (Figure 2D). HMMM increases the diffusion rates of lipids in the membrane allowing relevant binding sites to be identified more efficiently. Examples of HMMM being applied successfully are in the identification of several orientations of KRas4b binding to membranes containing anionic lipids (74) and in the identification of canonical and alternative binding sites for $PIP_3$ on the Grp1 PH domain (75).

*Accelerating computational performance by coarse graining*

In addition to AA MD and HMMM, coarse-grained (CG) MD methods are popular for looking at PMP interactions with membranes. In physics-based CG approaches, groups of atoms are represented by a single particle rather than explicitly. This massively reduces the degrees of freedom needed to describe a specific system, as well as allowing a longer MD timestep to be used, both of which potentially increase simulation speeds by several orders of magnitude. Different levels of coarse graining, as well as the application of continuum models, are outlined in detail by Moquadam *et al.* (57), and will be briefly discussed below.

In terms of characterising specific protein-lipid interactions, considerable success has been achieved using the Martini 2 (76–78) and 3 (79) force fields, as well as variants of this: polarizable Martini (80,81), BMW-Martini with polarisable water (82,83) and Martini 2.3P for cation-π binding (84). Martini 3 was recently tested on 12 different PMPs and lipid-binding peptides, and the authors found that the force field was in most cases able to identify experimentally known binding interfaces (85). Some false negative and false positive results were, however, also reported, meaning that a validity check with AA simulations of the CG binding poses is recommended. An important observation was that the elastic network which is often applied to conserve secondary and tertiary protein structure in Martini, prevented the PMPs from undergoing conformational changes upon binding. Aside from Martini, other CG force fields are also able to be used for looking at PMP-lipid interactions, including SIRAH (86–88), ELBA (89) and dissociative particle dynamics (DPD) (90,91). Each have different advantages, although when choosing a force field it is worth considering what lipids are already parametrized in that model (92).

Implicit or continuum models are typically very fast for PMP surface identification (93), however specific PMP-lipid interactions can be more challenging to model this way. Poses generated from this can, however, be used to guide MD simulations. An early example of this is work done on prostaglandin H2 synthase-1 binding to a PC membrane (94). In particular, this approach could potentially be extremely powerful for orienting PMPs prior to simulation, with several tools and databases available for this, including the DREAMM tool (http://dreamm.ni4os.eu/; (95)) and the Orientations of Proteins in Membranes database (15,96).

*PMP self-assembly with coarse-grained MD can reveal intermediate states*

Coarse graining makes it possible to simulate microsecond trajectories as a matter of course with basic computer hardware. This is very important for PMP/membrane systems, where the membrane and PMP need time to equilibrate and sample multiple configurations, often including PMP rotational dynamics, which can be relatively slow. The first coarse-grained simulations with PMPs encountering lipid membranes were performed 15 years ago, such as for SGTX1 binding a POPC or POPE/POPG membrane (97). Since then, the method has been applied to many different systems, such as PIP5K1A (98) and several PH domains, including Grp1 (99), Dok7 (100) and kindlin-3 (101). As computational power increases, even more comprehensive analyses are possible (102).

It is worth noting that many of these studies improve their sampling of PMP-membrane interactions through running several independently initiated simulations, typically between 5 and 25 repeats per system. The reported variation between repeats can be significant, reflecting the complicated energy landscape for a PMP with several membrane interaction sites binding a membrane with different lipid constituents, as well as the likely non-ergodic nature of single MD trajectories. Through these repeats it is often possible to see multiple minima reflecting different binding sites on the same PMP (21,98,100). Some of these may reflect genuine metastable states which act as important intermediates before the final binding pose. Such intermediate binding poses are extremely difficult to determine experimentally, as any ensemble data will likely be dominated by the final and energetically most favourable binding pose. MD simulations can thus provide a unique insight into the binding (and unbinding) pathways of PMPs.

*The best of both worlds: multi-scale modelling*

The speed-up obtained through coarse graining does come at a price. For the Martini force field this includes a simplified description of electrostatic interactions (103) and exaggerated protein stickiness (104), as well as other potential limitations. To ameliorate this, multi-scaling can be utilized. Here, the configurational landscape of given system is rapidly sampled using CG, such that key states can be converted to AA description for analysis of these state in higher resolution (51,105). This would normally be a dominant binding mode or interesting intermediate state. An example of this being applied is to the voltage-sensitive phosphatase (106). First, a binding pose was generated using CG simulation, and converted to AA. This AA simulations then allowed the position of a bound $PIP_3$ molecule to be refined with respect to PTEN, to describe the specific residue-lipid bonds in higher resolution than possible in CG. Of course, it is possible for any changes that occur in the AA to be converted back to CG for additional analysis. This might be important when looking at conformational changes, which can be difficult to accurately model using CG where an elastic network is often applied to stabilise the protein structure.

For HMMM, the increased diffusion rate likewise comes with a penalty. For instance, it is unclear how the exchange of lipid tails for organic solvent will impact the entropy of the membrane, and therefore the energetics of PMP binding. Similarly, the effects of different lipid tails and cholesterol presence are difficult to accurately account for. However, it is also possible to convert HMMM representations to an AA description once the desired pose has been sampled (73) and thus validate the HMMM results. This was done, for instance, to describe the functional cycle of the fatty acid transfer protein FakB1 (22) or GRP1 PH binding to anionic lipid (75). Similarly, systems can be converted from CG to HMMM description, as has recently been done for the binding of vinculin to PIP$_2$ (107).

*Utilizing multi-scale modelling for comparative studies of protein domain families*

The speed-up provided by coarse-graining combined with hardware and software improvements (including improved GPU utilization, e.g. (108,109)), has allowed MD studies of whole families of lipid-binding domains. This includes a comprehensive study of more than 10 different PH domains and their interaction with PIP$_2$- and PIP$_3$-containing membranes (110), as well as a study of 6 different C2 domains (21) binding to anionic and zwitterionic model membranes. Such studies highlight the utility of multi-scale simulations, where many systems can be concurrently analysed to provide a high-resolution comparison between domains from different proteins.

*Selection/identification of binding sites from a trajectory*

Whether a trajectory is generated by CG, HMMM or AA MD simulation, a critical step is the identification of lipid binding sites from the trajectory data (Figure 3). Analysis by visual inspection is always a useful step, but this can be somewhat subjective and time-consuming, especially if using multiple repeats. Therefore, systematic analysis strategies are necessary.

As a simple analysis, different binding poses, including intermediate binding modes, can be mapped out through simple geometric analysis of the trajectories. An example of this is producing a 2D heatmap comparing PMP-membrane distance vs PMP orientation (21,98,100). By quantifying the sampling density of the PMP along these two coordinates, it is possible to draw qualitative observations about stable and metastable binding modes, especially in combining multiple simulations into a single ensemble. Important residues and binding pockets which interact with membrane lipids can then be identified by e.g. contact analysis. This is a somewhat complex exercise, but fortunately specialised tools have recently been developed for this purpose.

*Dedicated analytical tools for analysis and identification of lipid binding sites*

The study of PMP interactions with lipids is related to the more established field of ligand binding to soluble proteins. For this latter topic, a range of dedicated tools exists for identification and analysis of binding pockets (111). The methods are, however, not directly transferable to PMPs even for specific lipid binding sites, as there are many differences between a soluble ligand and a membrane-bound lipid, including the limited rotation and vertical (membrane-perpendicular) translation of the lipid. Therefore, dedicated tools can be useful. We highlight two recent examples below.

The first is ProLint (112). ProLint allows the user to visualise and analyse lipid-protein binding sites and, e.g., identify contacts between specific lipid types and residues. A contact is defined as a distance below a user defined cut-off (between 3 and 8 Å). By examining, e.g., a heatmap of lipid contacts on the PMP, the user can identify binding sites. ProLint can be used as a python package or via the web interface.

The second tool is PyLipID (113) PyLipID is a Python-based package which reads in trajectory files and creates an interaction profile for each individual residue-lipid pair over the course of the simulation. It then applies a network analysis method to cluster residues which simultaneously binding the same lipid, to allow identification of specific lipid sites. A similar approach has been applied to interactions of Kir2.2 with both cholesterol (114) and with a more complex mixture of lipids (115). This method also has conceptual overlap with a recent approach to site identification looking at a frame-by-frame h-bond interaction profile for PIP$_2$ with vinculin (107). One useful feature of PyLipID is that it gives statistics for identified binding sites, including site occupancy (% of time a lipid spends in the binding site) and a residence time and $k_{off}$. Like ProLint, these measures rely on a user-provided cut-off, typically between 4 and 8 Å). Additionally, PyLipID produces representative poses of the bound lipid for either creation of figures or downstream analysis.

There are multiple additional programs designed to analyse membranes from MD simulations. These might not necessarily be optimal for looking at specific protein-lipid interactions but might offer useful tools for looking at the interaction of the PMP with the membrane more generally, including how the membrane changes upon PMP binding. Of particular note are FATSLiM (116), MemSurfer (117), LOOS (118,119) and LiPyphilic (120). This range of programs, along with ProLint and PyLipID mentioned above, showcase the increasing desire for more rigorous and detailed analyses of membrane simulations, and the impressive commitment of the academic community to fulfil this demand. Going forward, we expect much more progress in both how people develop and use scientific software (121).

## Complex binding sites

The structural resolution of MD also allows more complex interactions to be observed, including the clustering (leading to avidity effects) or perturbation of lipids upon binding. Here, we will briefly outline some examples of what type of more complex binding events have been investigated using MD.

### Lipid variety

The type of lipid, including variation in both headgroup and lipid tails, can potentially make a big difference in how PMPs associate with the membrane. For example, a recent study used AA MD coupled with fluorescence microscopy to demonstrate that saturated and unsaturated GM1 have different properties when bound to Ctx (Figure 1B), including ability to co-cluster with GPI-anchored proteins (122). The authors also showed that saturated GM1 was more able to couple *trans*-bilayer to PS than unsaturated GM1, which has implications for the ability of cholera toxin to interact with the actin cytoskeleton.

### Multiple lipids and lipid clustering

Structural techniques such as X-ray crystallography and cryo electron microscopy can, in some cases, resolve one or more tightly bound lipids in their density maps. However, MD simulations on the same systems can often reveal additional lipid binding sites around the PMP. For instance, there are several MD studies which report the importance of multivalent PIPs binding to PMPs, and suggest that PIP clustering can help stabilise the PMP on the membrane. Notably, Yamamoto et al investigated $PIP_3$ bound to the PH domain of GRP1 and showed that experimental binding values could only be reproduced in simulations after including multiple PIPs in the binding event (123) (Figure 4C). That is, avidity must be taken into account as affinities between single molecules are not trivially additive. Binding of multiple $PIP_2$ were likewise observed for different membrane-binding C2 domains (21), for Ebola virus matrix protein VP40 (124) and for Brag2 (125), and multiple PI(4)P were found to bind the FYVE domain of early endosome antigen 1 (16).

MD can also investigate how the binding of multiple lipids affects the orientation of the PMP. As an example, several $AB_5$ exotoxin structures have been identified bound to fragments of their specific glycolipid receptors. This binding is multivalent; there are 5 binding sites and 5 bound lipids. However, Sridhar et al (126) used CG simulations to allow Ctx to freely bind GM1 in the membrane, and saw that GM1 typically bound 3 out of the 5 sites, which lead to a tilted conformation. Both tilting (127) and occupancy of 3 sites (62) have also been see in AA, and the occupancy of 3 sites was shown to be the dominant state for Ctx using flow cytometry (128).

Another application of MD is the study of non-specific lipid clustering and raft formation upon binding of PMPs. One example is the raft-formation of dipalmitoyl-glycero-phosphocholine (DPPC) induced by influenza hemagglutinin (129). Similarly, DPD simulations have demonstrated that if multiple copies of Ctx are present, then the toxins can also interact allowing larger regions of lipid clustering to occur (130), as well as that multiple copies of Shiga toxin can cluster on the membrane (131).

*Lipid order*

A third notable membrane variation that can be studied with MD is that of lipid order. A prominent example of this is a study of the matrix domain of HIV-1 gag, which has a myristoylated N-terminal domain that inserts into the bilayer upon binding. However, this insertion was shown only to take place in ordered membrane models, which contained cholesterol and only a small amount of unsaturated lipid tails (132).

*Membrane curvature and very coarse-grained force fields*

Several MD studies have investigated PMP-induced membrane curvature and some PMPs are localized to, or cause, regions of membrane curvature (see, e.g. (20,127,133–135)). We have, however, chosen not to cover this in depth in the present review, as we focus on more specific interactions between PMPs and lipids. For the same reason, we have chosen not to cover coarser membrane and protein representations than the Martini representation, such as with mesoscale simulations (136). That said, incorporating mesoscale resolutions into studies along with CG and AA (137) can potentially be very powerful, especially for looking at such phenomena as membrane curvature (138). Similarly, continuum models can be used to rapidly predict PMP association with regions of curvature via the PPM 3.0 webserver (https://opm.phar.umich.edu/ppm_server3; (139)).

*Comparison with experimental data*

Finally, although MD is a powerful tool in the identification and characterisation of protein-lipid interactions, integration with experimental data is still very powerful, both for validation of the MD, and provision of a broader cellular and physiological context to the data. Therefore, future MD work will continue to require additional experimental data. Promising steps are being made in structural experimental techniques to measure specific protein-lipid interactions.

X-ray crystallography has for many years been the dominating technique for determining specific lipids cites on PMPs, by co-crystallization with lipid mimetic molecules, and has, e.g., been used to identify the PIP binding site of FYVE from EEA1 (140) and several PH domains (141). Nuclear magnetic resonance (NMR) can also be used to resolve lipid binding sites of PMPs (142) and is complementary to crystallography as it provides the solution structure and contribute with dynamic information (22).

In recent years, cryo electron microscopy (cryo-EM) has improved its resolution to a point where tightly bound lipids can be resolved from the electron density maps, as seen for integral membrane proteins, e.g., the P4-ATPase lipid flippase (143). To our knowledge, however, there are still no cryo-EM structures of lipids bound to PMPs. One advantage of cryo-EM over crystallography, is that cryo-EM can solve protein structures bound to lipid nanodiscs or lipid nanotubes, and thus be used to determine the orientation with respect to that model membrane. Cryo-EM has, e.g., been used to determine the binding orientation of factor VIII (144).

Neutron and X-ray reflectometry is another promising technique for assessing PMP binding orientation. Here, the average electron or neutron scattering length densities can be measured with high accuracy perpendicular to the plane of the membrane. This has been utilized, e.g., to determine the orientation of tubulin (145), and to investigate membrane-induced conformational changes of alpha synuclein (146). As mentioned previously, EPR is another important technique for determining orientation of membrane-bound PMPs (147).

Finally, small-angle neutron scattering (SANS) provides an interesting tool for investigating membrane-induced conformational changes of PMPs. By exchanging hydrogen with deuterium in solvent and/or lipids or detergents, the signal from the membrane mimic can be rendered effectively invisible. In that way, the membrane mimic is not providing any signal, and membrane-induced structural change of the PMP can be measured. This was done for myelin and matched-out micelles (148). That method is limited in resolution by the difference between micelle and lipid head and tail-groups, as only the average scattering is matched out. Thus, the signal from the micelles/lipids is only matched out at the low $q$-values (low resolution). To improve this, special partly-deuterated detergent and lipids have been developed to match out in the full measured range (149). This method has been shown feasible for bicelles (150), micelles (151) and nanodiscs (152). Despite improvements, SANS is limited to about 10 Å resolution at best.

## Section 2: Use of molecular dynamics to estimate binding affinities of interactions

Increasingly, many researchers are not just interested in 'if' and 'where' a given lipid might bind to a PMP of interest, but also how strong the interaction is, i.e. its affinity. This can be important, as determining the affinity provides insight into how likely a given interaction is in the complex biological context of the cell. Research into lipid affinities for PMPs is arguably less developed than for integral membrane proteins (153), but many of the developments made there are now also being applied to PMPs.

There are different ways of estimating affinities from MD simulations, the simplest being to run an unbiased simulation and compare the likelihood of the PMP being bound to the lipids vs unbound. The likelihood is quantified simply as the number of frames the PMP stays in each state. This method typically requires a prohibitive amount of sampling to achieve convergence, as several binding and unbinding events must take place during the simulation. Thus, enhanced sampling MD simulations are often run, in which the underlying potential of the force field is perturbed to drive the simulation to sample the landscape of PMP-lipid binding more rapidly.

*Avoiding energy minima with enhanced sampling techniques*

A range of enhanced sampling methods have been developed to drive simulations towards sampling of rare events. These methods can be split into reaction coordinate-based methods and reaction coordinate-free methods (154). A reaction coordinate, also denoted a collective variable, defines a coordinate that promotes the reaction of interest. In this case, the reaction is likely to be binding/unbinding of the PMP to a lipid membrane, in other cases the reaction could be ligand binding, or reactions interpreted in a broader sense, such as conformational changes.

Metadynamics (MetaD) is an example of a method that needs a reaction coordinate. For a PMP/membrane simulation this would typically be the membrane-protein distance. The MetaD algorithm stores the values of that variable while running the simulation and gives energetic penalties when states are revisited (155,156). By adding penalties, the simulation is eventually driven out of energy minima (Figure 4A). The added potential can then be processed to produce a free energy landscape. A variation of MetaD is well-tempered MetaD, where the applied potential is gradually decreased to achieve convergence (157). Well-tempered MetaD was recently applied to get 1D energy landscapes for the Grp1 PH domain binding to different membranes (75). Notably, MetaD is not restricted to a single reaction coordinate (usually denoted 'collective variable' in the context of MetaD), so MetaD has also been applied to produce 2D energy landscapes for integral membrane proteins using CG simulations (158) and for a C2 domain using AA simulations (159), suggesting that this is a viable option for PMPs, where both distance from the membrane and rotation of the PMP are of interest.

Accelerated MD (aMD) is an example of a reaction coordinate-free enhanced sampling method. The underlying potential is raised if it is below a defined threshold energy (160). Just like MetaD, the purpose is to drive the simulation out of energy minima to sample more unlikely events (Figure 4A). In a recent study, aMD was used to investigate the flexibility of Lewis Y antigens when bound to Ctx, and MM/GBSA to determine affinities between the antigens and Ctx (161).

It can be challenging to define a reaction coordinate, e.g. for conformational changes. Moreover, reaction coordinate-based simulations are biased towards a given reaction pathway and are therefore not explorative in nature. Thus, reaction coordinate-free methods are alluring. However, for most PMP/membrane systems, the dominant reaction coordinate is easy to define, namely the membrane-protein distance, and, more importantly, defining a reaction coordinate can prevent unwanted events; if a reaction coordinate is not defined for the system, there will inevitably be many ways out of a given energy minimum (Figure 4A), which can lead to, e.g. irrelevant conformational changes or unfolding of secondary structure etc.

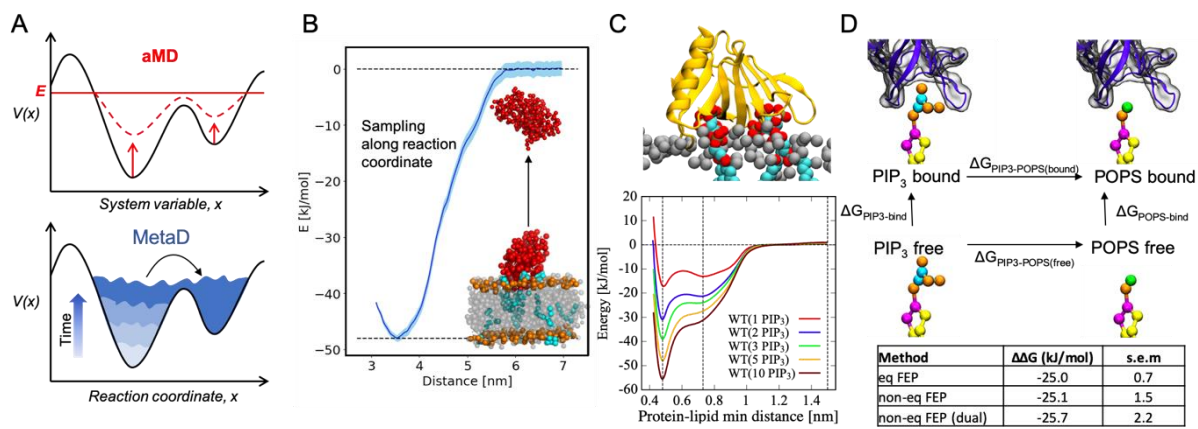

***Figure 4. Free energy calculation methods in MD.*** *(A) Schematic illustration of two enhanced sampling methods, that both alter the energy landscape to force the simulation out of energy minima; accelerated MD (aMD) and metadynamics (MetaD), see the main text. (B) Simple potential of mean force (PMF) calculation. The PMP is sampled along a reaction coordinate, typically the membrane-protein distance and the force necessary to keep the protein in place along this coordinate is monitored. From these forces, a PMF can be calculated. Shown is C2 from PTEN bound to an anionic membrane* (21). *(C) Complex PMF. PMFs calculated for binding of PH domains with, respectively, 1,2,3,5 and 10 PIP₃ bound* (123). *(D) Thermodynamic cycle as used in FEP for switching between PIP₃ and POPS for a PMP system. The difference between ΔG_{PIP3-POPS(bound)} and ΔG_{PIP3-POPS(free)} is the relative difference in PMP-lipid interaction for PIP₃ and POPS. Below are the values from preliminary analyses of Grp1 PH binding FEP calculations. Note that these data were generated for this review, with a full manuscript to follow.*

## PMF calculations

One prime example of a free energy method for PMP-membrane interactions is potential of mean force (PMF) calculations. There are different ways of running PMF calculations, with one of the simplest being applying a directional steering force (steered MD) to the system along a reaction coordinate of interest, in this case usually the distance between the centre of mass of the PMP and the centre of mass of the membrane. The work required to move the system along this reaction coordinate can be recorded and used to generate an estimate of the free energy of binding (162). This has been applied previously to the PH domain from RRP, whereby binding energies were estimated for different membranes (163), and to the PMP dihydroorotate dehydrogenase (164). In the latter, the PMP was pushed into a model membrane to estimate the PMP insertion depth, by monitoring the opposing reaction force.

PMFs can also be constructed through the application of umbrella sampling. Umbrella sampling is a reaction coordinate-based enhanced sampling method. In umbrella sampling, frames are typically extracted from a steered MD simulation at regular positions along the reaction coordinate. The frames are then used as starting points for a parallel series of simulations (umbrella windows), in which the protein is constrained to maintain the input value of the collective variable, i.e., stay at a specific distance from the membrane. The forces imposed by umbrella potential can then be used to construct a PMF using a technique such as the weighted histogram analysis method (165).

The umbrella sampling PMF method has been applied to a number of PMPs, including studies of PH domains (99,123), C2 domains (21), the FYVE domain of early endosome antigen 1 (16) and the related lipidated GTPase Rab5 (23), which are both involved in early endosome recruiting. An example for a C2 domain is given in Figure 4B. Umbrella sampling PMFs have also be constructed for different binding modes of the soluble domain from the mitochondrial translocase subunit TIM50 (166). In all of those studies, the PMF was generated using coarse-grained MD. PMFs have also been calculated from AA simulations, e.g. to estimate the membrane penetration depth of the C2A domain of synaptotagmin (24) and to probe the binding energy of actin-binding proteins bound to PIP-containing membrane (167).

*Free energy perturbation calculations*

A different method for estimating the interaction energy between a PMP and a specific lipid is free energy perturbation (FEP) calculations. Here, alchemical transformations of a target lipid into a different lipid are performed on an unbound lipid in the membrane and for the lipid bound to the protein, and the ΔG of each process computed using a technique such as MBAR (168). The difference between the free and bound lipids, the ΔΔG, can then be calculated (Figure 4D). The calculations are normally run for the target lipid to a generic lipid, such as POPC, the target lipid can be converted to a different lipid of interest, such as switching from $PIP_3$ to $PI(4,5)P_2$ to $PI(4)P$ to PI to compare the energetics of how different PIP lipids bind the same site (169). We suggest refs (158,170) for a more detailed run-down of this process.

Analysis done on integral membrane proteins suggests that the values obtained from FEP in CG are approximately equivalent to PMFs (158), which has also been shown for PMPs (21). But FEPs generally require less sampling time, which can be very beneficial, especially when looking at multiple different lipids binding to the same protein. For PMPs in particular, FEPs are potentially very advantageous, as the length of a PMF reaction coordinate for a PMP can be up to 4-5 nm (16), and convergence can be difficult due to the slow rotation of the PMP. In addition, the 'free' and 'bound' FEP calculations can theoretically be done in the same simulation system, with the bound lipid in one membrane leaflet and the free in the other. This would serve to reduce computation time even further.

Certain technical advances would make FEP of PMP-lipid systems even more computationally efficient, in particular the use of non-equilibrium (171). In non-equilibrium FEP, long (e.g. 100 ns) simulations are first run in both states A (target lipid) and B (generic lipid), and then very short (e.g. 200 ps) non-equilibrium simulations are run for several snapshots (e.g. 20) going from A to B and B to A. To enable statistical analysis, the whole cycle is repeated a number of times (e.g. 10 repeats). As for standard FEP, this is run for both the unbound lipid in the membrane, and the lipid bound to the protein.

For the Grp1-PH domain, non-equilibrium FEP agrees well with equilibrium FEP (Figure 4D; 'eq FEP' vs 'non-eq FEP') but costing only ~2 μs vs 10 μs. In addition, by combining both the free and bound lipid in the same system, this can be halved to only 1 μs per system, without impacting the final value (Figure 4D; 'non-eq FEP dual'). Other potential savings in time might come from ways of increasing the sampling rate within each window, such as with the accelerated weight histogram method (172,173). This will be particularly useful for any FEP calculations run using AA force fields.

*Advanced energy analysis methods from unbiased simulations*

The free energy of a system can also be extracted from a series of trajectories using Markov State Modelling (MSM), where the transition states and corresponding transition rates are analysed (61). As MSM provides information about transitions, it constitutes a promising tool for precise descriptions of complex binding events involving intermediate states. Another intriguing new method demonstrates that it may be possible to directly extract affinities in the form of apparent dissociation constants ($K_d$) from the simulation of binding saturation curves (174).

Both analysis methods are still reliant on the unbinding of the PMP from the membrane, which might require prohibitively long simulation times for certain PMP-lipid compositions. Therefore, they can readily be combined with enhanced sampling techniques such as MetaD, aMD or umbrella sampling.

*Coarse-grained versus all-atom simulations for energy calculations*

Both CG and AA approaches have limitations. CG MD involve approximating the chemical landscape, as described previously. In cases with available experimental data, however, good consistency has reported between experimental energies and energies obtained with PMF using Martini 2 (16,123). AA MD is a more accurate description of the system, but it is more challenging to obtain convergence (38). Therefore, researchers are often left with the choice of either accepting approximative models such as Martini or HMMM, with their respective limitations, or accept non-converged PMFs from AA simulations. This is set to change, however, as increasing computer speeds and improved MD algorithms will make AA more affordable, and improved force fields will make CG more accurate (79).

*Comparison with experimental data*

Comparison of predicted free energy values with those from experimental data is beneficial. Surface plasmon resonance is the most frequently used experimental technique assessing PMF/membrane binding energies, and has been used for a range of protein domains, including PH domains (36,39), FYVE domains (175–177), ANTH (17), cytosolic phospholipase A2α (178), the adaptor protein Amot (179) and the Bcl2-associated agonist of cell death (BAD) (180). Competitive FRET and isothermal titration calorimetry can likewise determine binding free energies (141,181).

A central question is whether the free energy values from simulations correspond well with those from experimental analysis. Good consistency with experimental values has been observed for AA FEP calculations on three different outer-membrane associating PMPs (19). This study used a version of the CHARMM36 force field optimized for cation-π interactions (182,183) (Figure 2D). Energies determined by PMF from CG simulations with Martini 2 likewise showed good consistency for the PH domain from GRP1 (123) and the FYVE domain of EEA1 (16). However, more systematic and comprehensive experimental validation of computational results is an essential mean towards refining the computational methods.

The role of MD is not simply to reproduce the values which can be obtained experimentally, however. When analysed correctly, MD simulations are able to provide a range of structural and energetic data, which would be extremely difficult to obtain experimentally. For instance, the binding energy can be divided into contributions from different types of lipids, or from different binding sites on the protein (see, e.g. Figure 4C). Moreover, MD simulations provide a tool for relatively easy investigations of the energetics of, e.g. introducing mutations, or changing the membrane composition, or perturbations induced by small molecule binding to the membrane or to the PMP. The latter is highly relevant in drug design.

## Conclusions and outlook

Here we have presented an overview of work done to identify and characterise PMP binding to specific lipids in biological membranes using MD. We have focused on PMPs, and point readers also interested in integral membrane proteins to other reviews (184,185). We have chosen to highlight a number of key works where researchers have identified specific protein-lipid interactions, and several studies where the energetics are also considered.

Future studies may aim to incorporate more advanced data analysis methods to eventually reduce the dependence on lengthy MD simulations for site identification, e.g. machine learning approaches. Current work has demonstrated that certain structural features of PMPs can be used to rapidly predict membrane-binding properties, including hydrophobic protrusions (186). This means that certain predictions can be made extremely cheaply, such as with the PePrMint webserver (https://reuter-group.github.io/peprmint/).

There exist a number of challenges facing development of MD for PMP analysis, which arise from the biological complexity behind the binding process. For instance, such biological processes as how protein complexes form at the surface of biological membrane are out of reach for all but the lengthiest simulations (60). Additionally, more care should be taken in how systems are built for simulation. As an example, a recent study has demonstrated that the protonation states of charged PIP lipids might also play a considerable role in PMP binding (187), something which is generally overlooked. Similarly, key features like how ions help stabilise the PMP binding increase complexity and hence may require costly simulations, e.g. with polarizable force fields (Figure 3B).

A key area of future work will be to see how membrane association affects the PMP itself. This is more challenging to address than simple site identification, in part because it typically requires far longer time scales for the effects to be observed, and the restrictive nature of coarse-graining on protein structure, including the elastic network in the Martini force field (79), means that AA simulations are likely necessary. In addition, there is far less experimental data to compare to, making interpretation of MD data more difficult. In a recent study, Gullett *et al.* describe the binding of a bacterial fatty acid protein in its open and closed states, allowing production of its complete exchange cycle (22). Here, MD was performed separately on an open and closed conformation, both determined by X-ray crystallography. A current challenge is thus to directly simulate binding-induced conformational change of PMPs. As such, much of the current focus has been on smaller and more localised changes. For instance, several studies have looked at how loop dynamics are altered on membrane association, including for cytochrome P450 (188), or conformational changes of peptides. As an example of the latter, a study by Davis and Berkowitz reported increased helical content of an amyloid-β peptide upon binding to an anionic membrane (189).

Finally, more complex analysis methods are needed to shed light on transient binding modes, intermediate modes and binding pathways. An obvious tool for this would be Markov State Models as mentioned previously. This includes the challenge of directly combining simulations and experiments to achieve more accurate models for the transition states (190).

Going forward, we anticipate many advances in MD algorithms and force fields, improved hardware, better analysis methods and superior strategies for integration with experimental data. This will all result in fruitful insight into the binding of PMPs to lipid membranes. MD may also play a key role in development of a new class of drugs, targeting PMPs or membranes (3).

## Acknowledgements

AHL is funded by the Carlsberg Foundation (CF19-0288) and Lundbeck Foundation (R347-2020-2339). LHJ is funded by the UKRI-BBSRC Interdisciplinary Bioscience Doctoral Training Partnership (BB/M011224/1) and the ISIS facility development studentship program. MSPS and RAC are supported by Wellcome (208361/Z/17/Z). Free energy calculations were run using the ARCHER/ARCHER2 UK National Supercomputing Service (http://www.archer.ac.uk), provided by HECBioSim, the UK High End Computing Consortium for Biomolecular Simulation (hecbiosim.ac.uk), which is supported by the EPSRC (EP/L000253/1).

Larsen et al 2022